\documentclass[12pt]{article}
\usepackage[utf8]{inputenc}
\usepackage{lipsum}
\usepackage{hyperref}
\usepackage{enumitem}
\usepackage{float}
\usepackage{graphicx}
\usepackage{caption} 
\usepackage{calc}
\usepackage{fancyhdr}
\usepackage{tcolorbox}
\usepackage{hyperref}
\usepackage{wrapfig}
\usepackage{graphicx} 
\usepackage{amssymb}
\usepackage{amsmath}
\usepackage{multicol}
\usepackage{braket}
\usepackage{abstract}
\usepackage{multicol}
\usepackage{xcolor}
\usepackage{appendix}
\title{\large\bf Asymptotic Generation of Kerr Geometry from Schwarzschild via BMS Supertranslations}
\author{\normalsize {\sc Nihar Ranjan Ghosh\footnote{\tt g.nihar@iitg.ac.in}\ \ and Malay K. Nandy\footnote{\tt mknandy@iitg.ac.in \rm (Corresponding Author)}}\\
\normalsize \em Department of Physics, Indian Institute of Technology Guwahati\\
\normalsize \em Guwahati 781 039, India}
\date{September 1, 2025}

\begin{document}

\maketitle

\begin{abstract}
The Bondi-van der Burg-Metzner-Sachs (BMS) group, as the asymptotic symmetry group of asymptotically flat spacetimes, plays a central role in connecting infrared structures of gravity with soft theorems and gravitational memory. In this work, we investigate the extent to which BMS supertranslations can relate physically distinct black hole geometries. Focusing on the Schwarzschild and Kerr solutions, we show that the asymptotic structure of the Kerr spacetime can be generated from the Schwarzschild geometry via two successive supertranslations. These transformations yield a Kerr-like geometry at null infinity and reveal two distinct classes of supertranslation functions. The first, composed of $l=1$ spherical harmonics, corresponds to center-of-mass displacements and encodes the translational sector of the BMS group. The second, characterized by an infinite series of even-parity Legendre polynomials ($l \geq 2$), captures the intrinsic mass multipole structure of the Kerr spacetime. Our result illustrates how BMS supertranslations can act as symmetry transformations linking asymptotically flat black hole geometries, and that they encode physically meaningful soft hair consistent with the multipole structure of rotating black holes. This work supports a unified description of soft degrees of freedom in black hole spacetimes and underscores the role of infinite-dimensional asymptotic symmetries in gravitational physics.
\end{abstract}

\maketitle

\section{Introduction}

The Bondi-van der Burg-Metzner-Sachs (BMS) symmetry group \cite{bondi1962gravitational,sachs1962gravitational} represents one of the most profound discoveries in modern theoretical physics, fundamentally reshaping our understanding of asymptotic symmetries in general relativity. Initially identified in the 1960s as the asymptotic symmetry group of asymptotically flat spacetimes, the BMS group extends the finite-dimensional Poincar\'e group by an infinite set of generators known as {\em supertranslations}. These transformations correspond to angle-dependent translations at null infinity, and they significantly enrich the phase space structure of general relativity by distinguishing between spacetime configurations that are otherwise indistinguishable under Poincar\'e symmetry.

The emergence of the BMS group was unexpected. It was originally anticipated that the symmetries acting nontrivially on asymptotically flat spacetimes would be identical to the isometries of Minkowski space. Contrary to this expectation, Bondi, van der Burg, Metzner, and Sachs demonstrated that the symmetry structure at null infinity is much richer, encompassing the infinite-dimensional BMS group. This discovery has led to deep insights across gravitational theory, particularly in the context of the so-called {\em infrared triangle} \cite{strominger2018lecturesinfraredstructuregravity}, whose three corners comprise asymptotic symmetries, soft theorems, and gravitational memory effects.

A parallel development occurred in the context of asymptotically anti-de Sitter (AdS) spacetimes, notably through the work of Brown and Henneaux \cite{brown1986central}, who demonstrated that the asymptotic symmetry algebra of $\mathrm{AdS}_3$ gravity consists of two copies of the Virasoro algebra. While the focus of this work remains on asymptotically flat spacetimes, the structural similarities in both settings highlight the universality and physical relevance of asymptotic symmetry analysis \cite{CADONI1999165, hotta1998asymptoticisometrydimensionalantide, henneaux1985asymptotically, Comp_re_2016}.

\subsection{Gravitational Memory and the Infrared Triangle}

Within the infrared triangle, the {\em gravitational memory effect} constitutes a classical and observable manifestation of BMS symmetries. First identified in the linearized regime by Zel’dovich and Polnarev \cite{Zeldovich:1974gvh}, and later generalized to the nonlinear setting by Christodoulou and others \cite{Christodoulou, Braginsky:1985vlg, Braginsky:1987kwo, PhysRevD.44.R2945, PhysRevD.46.4304, PhysRevD.45.520, Favata:2010zu, Tolish:2014bka, Tolish:2014oda, Winicour:2014ska}, the memory effect refers to the permanent displacement of freely falling test particles following the passage of gravitational radiation. This phenomenon has been shown to be intimately related to BMS supertranslations, as it encodes information about the radiative history of spacetime via conserved charges at null infinity \cite{Strominger:2014pwa, strominger2018lecturesinfraredstructuregravity, PhysRevD.92.084057, Flanagan:2015pxa, Pasterski:2015tva}.

Recent investigations have extended the notion of gravitational memory to include spin and center-of-mass memory effects \cite{Pasterski:2015zua, PhysRevD.98.064032, Compere:2016jwb}, as well as electromagnetic analogues \cite{Bieri:2013hqa, susskind2019electromagneticmemory}. In the black hole context, Donnay et al.~\cite{PhysRevLett.116.091101} introduced the black hole memory effect, demonstrating that transient radiation can induce permanent deformations in the near-horizon geometry. This observation establishes a direct link between gravitational memory and the emergence of soft hair on black holes. Additional insights have been obtained in the study of extremal black holes, where memory effects reveal new features due to horizon degeneracy \cite{PhysRevD.102.044041}.

\subsection{Soft Theorems and Asymptotic Symmetries}

The third component of the infrared triangle, the {\em soft theorems}, originates from early studies in quantum electrodynamics, notably the work of Bloch and Nordsieck \cite{PhysRev.52.54}, Low \cite{Low:1954kd, Low:1958sn}, Gell-Mann and Goldberger \cite{Gell-Mann:1954wra}, and Yennie et al.\ \cite{Yennie:1961ad}. In gravity, the soft theorem was developed by Weinberg \cite{Weinberg:1965nx}, who showed that the emission of low-energy gravitons in scattering processes is governed by universal behavior independent of the specific details of the interacting fields.

Crucially, these universal behaviors are now understood to correspond to Ward identities associated with asymptotic symmetries, thus establishing a direct connection between soft theorems and BMS charges \cite{Pasterski:2015tva}. In this picture, soft theorems provide a quantum field theoretic realization of the infinite-dimensional symmetry structure uncovered by BMS analysis, and serve as the basis for the {\em soft hair} theorem in black hole physics.

\subsection{Soft Hair and the Information Paradox}

The soft hair proposal has attracted considerable attention due to its implications for the black hole information paradox \cite{PhysRevD.14.2460}. Hawking, Perry, and Strominger \cite{PhysRevLett.116.231301} argued that the infinite set of conserved BMS charges leads to an infinite number of zero-energy excitations, the soft hair, on the  black hole horizon. These soft degrees of freedom could, in principle, encode information about the microstates of the black hole and provide a mechanism for information retention during Hawking evaporation \cite{Hawking:1975vcx, Hawking:2016sgy, haco2018black, PhysRevD.96.084032, Chu_2018, PhysRevD.108.044034}.

This proposal significantly modifies the classical no-hair theorem \cite{Israel:1967za, Israel.164.1776, PhysRevLett.26.331}, which asserts that black holes are characterized by only a finite set of global charges. The inclusion of soft hair reveals an infinite-dimensional structure not accounted for in the traditional framework \cite{strominger2017blackholeinformationrevisited}, and has important consequences for black hole entropy and unitarity in quantum gravity \cite{haco2019kerrnewmanblackholeentropy, PhysRevD.103.126020}.

A related perspective interprets soft hair as edge modes, which are boundary localized degrees of freedom arising from large gauge transformations that act nontrivially at the boundary of spacetime. These modes are essential for maintaining gauge invariance on subregions and have been widely studied in both gauge theory and gravity \cite{Donnelly:2014fua, Donnelly:2015hxa, Harlow:2015lma, Harlow:2016vwg, Maldacena:2016upp}. This viewpoint suggests that soft hair may represent a general feature of gauge theories with boundaries.

\subsection{Supertranslations and Black Hole Geometries}

In light of these developments, it is natural to ask whether physically distinct black hole solutions can be related by asymptotic symmetries, particularly BMS transformations. For example, is it possible that the Schwarzschild and Kerr solutions, both being the vacuum solutions of Einstein's equations, be connected via appropriate supertranslation transformations? Such a relationship would shed light on the structure of the gravitational solution space and the role of asymptotic symmetries in classifying vacua.

In this work, we explore this question by considering a Schwarzschild black hole expressed in the retarded coordinates $(u, r, \theta, \phi)$. We construct a class of asymptotic gauge transformations that transform the Schwarzschild geometry into one that asymptotically resembles the Kerr metric. Specifically, we show that two successive supertranslations, subject to appropriate gauge and large-$r$ falloff conditions, can generate a Kerr-like asymptotic structure. Since both Schwarzschild and Kerr geometries are vacuum solutions, this construction provides an explicit demonstration of how BMS symmetries can interpolate between distinct classical configurations in asymptotically flat spacetime.

The remainder of this paper is organized as follows. In Sec.~\ref{sec-kerr}, we present the Kerr solution expressed in Bondi coordinates and analyze its asymptotic structure, following Ref.~\cite{SJFletcher_2003}. In Sec.~\ref{sec-sch}, we construct explicit supertranslation transformations on the Schwarzschild geometry in the asymptotic limit at null infinity, whereas, in Sec.~\ref{sec-match}, we obtain appropriate analytical conditions under which the correspondence between the two geometries hold. Finally, in Sec.~\ref{disc}, we conclude the paper with a discussion.

\section{Kerr Geometry in Bondi Gauge} \label{sec-kerr}

In this section, we present the Kerr solution expressed in the Bondi coordinate system, which is well-suited for analyzing asymptotically flat spacetimes and their associated symmetries. The Bondi gauge provides a framework where the metric is adapted to null infinity, enabling a clear characterization of the asymptotic behavior of gravitational fields. Building upon the formulation introduced in Ref.~\cite{SJFletcher_2003}, we review the asymptotic structure of the Kerr geometry and highlight the key features relevant for subsequent analysis. This setup lays the foundation for generating Kerr-like asymptotics from supertranslation transformations acting on the Schwarzschild background to be discussed in the next section.

We start our analysis by first considering the Kerr metric in the Boyer-Lindquist coordinate system, given by 
\begin{equation}
    \label{kerr Boyer}
    \begin{split}
        ds^2_{\text{Kerr}}=-(1-\frac{2M_K\bar{r}}{\bar{\rho}^2})d\bar{t}^2&-\frac{4aM_K\bar{r}\sin^2\bar{\theta}}{\bar{\rho}^2}d\bar{t}d\bar{\phi}+\frac{\bar{\rho}^2}{\bar{\Delta}}d\bar{r}^2+\bar{\rho}^2d\bar{\theta}^2\\
        &+\sin^2\bar{\theta}\left(\bar{A}^2+2\frac{a^2M_K\bar{r}\sin^2\bar{\theta}}{\bar{\rho}^2} \right)d\bar{\phi}^2
    \end{split}
\end{equation}
with
\begin{equation}
    \begin{split}
        \bar{\Delta}&=\bar{r}^2-2M_K\bar{r}+a^2;\\
        \bar{A}&=\sqrt{\bar{r}^2+a^2};\\
        \bar{\rho}^2&=\bar{r}^2+a^2\cos^2\bar{\theta},
    \end{split}
\end{equation}
where $M_K$ and $a$ are the mass and angular momentum of the Kerr black hole.

To proceed with the BMS analysis, we first represent the Kerr metric in the generalized Bondi-Sachs (GBS) coordinate system. This is implemented by imposing the following conditions on the coordinates:
\begin{equation}
    \label{Boyer_GBS}
    \begin{split}
(i)~ &\bar{r} = \tilde{r}, \\
(ii)~ &\frac{\partial}{\partial \bar{t}} = \frac{\partial}{\partial u}, \\
(iii)~ &\frac{\partial}{\partial \bar{\phi}} = \frac{\partial}{\partial \phi}, \\
(iv)~ &\frac{du}{d\lambda} = \frac{d\theta}{d\lambda} = \frac{d\phi}{d\lambda} = 0, \\
(v)~ &\bar{\theta} \to \theta \quad \text{as } \tilde{r} \to \infty, \\
(vi)~ &\bar{\theta} = \pm \pi/2 \;\; \Leftrightarrow \;\; \theta = \pm \pi/2.
\end{split}
\end{equation}

These are the necessary and sufficient conditions, required to express the Kerr metric in terms of generalised Bondi-Sachs (GBS) coordinates. 
Condition $(i)$ implies the freedom in rescaling the radial coordinate $\tilde{r}$ after expressing the Kerr metric in GBS form. Conditions $(ii)$ and $(iii)$ ensure that the simple forms of the Killing vector fields are preserved in these new coordinates. Condition $(iv)$ requires that, in the GBS coordinates, the integral curves of the zero angular momentum null geodesics align with the lines of constant $u$, $\theta$, and $\phi$. Condition $(v)$ ensures that the new and old values of the angular coordinate $\theta$ coincide in the asymptotic limit, where they are expected to be indistinguishable for very large $\tilde{r}$. Finally, Condition $(vi)$ ensures that the equator of the old coordinates remains the equator in the new ones, preserving the natural plane of symmetry of the Kerr spacetime.

Thus, the Kerr metric, expressed in the generalized Bondi-Sachs form, with the coordinates related by \ref{Boyer_GBS}, is given by
\begin{equation}
    \label{GBS_kerr}
    \begin{split}
        ds^2_{\text{KGBS}}=&-(1-\frac{2M_K\tilde{r}}{\rho^2})du^2-2\frac{\rho^2}{B}dud\tilde{r}-2(1-\frac{2M_K\tilde{r}}{\rho^2})\frac{a \cos\theta}{C^2 \cosh^2{\alpha}}dud\theta-4\frac{aM_K\tilde{r}D^2}{\rho^2C^2}dud\phi\\
        &-4\frac{a^2M_K\tilde{r}\cos\theta D^2}{\rho^2C^4\cosh^2\alpha}d\theta d\phi+\frac{\tilde{r}(\tilde{r}\rho^2C^2\cosh^2\alpha+2a^2M_K\cos^2\theta)}{\rho^2C^4\cosh^4\alpha}d\theta^2\\
        &+\frac{D^2(C^2B^2\cosh^2\alpha+a^2\Delta\cos^2\theta)}{\rho^2C^4\cosh^2\alpha}d\phi^2,\\
  \end{split}
\end{equation}

where  
\begin{equation}
    \begin{split}
        &A=\sqrt{\tilde{r}^2+a^2},\\
        &\Delta=\tilde{r}^2-2M_K\tilde{r}+a^2,\\
        &B^2=A^4-a^2\Delta,\\
       &C=1+\tanh\alpha\sin\theta,\\
       &D=\tanh\alpha+\sin\theta,\\
    \end{split}
\end{equation}
and
\begin{equation}
\alpha =
\begin{cases}
\displaystyle \int^{\tilde{r}}_{-\infty} \frac{a \, dx}{\sqrt{x^{4} + a^{2} x^{2} + 2a^{2}M_Kx}} + 2\alpha_{0}, & \tilde{r} < 0, \\[1.2em]
\displaystyle -\int^{\infty}_{\tilde{r}} \frac{a \, dx}{\sqrt{x^{4} + a^{2} x^{2} + 2a^{2}Mx}}, & \tilde{r} \ge 0,
\end{cases}
\label{alpha}
\end{equation}
with
 \begin{equation}
   \alpha_0=-\int^{\infty}_{0} \frac{a \, dx}{\sqrt{x^{4} + a^{2} x^{2} + 2a^{2}M_Kx}} 
 \end{equation}
and 
\begin{equation}
    \rho^2=A^2-a^2\frac{D^2}{C^2}~~.
\end{equation}

For a more detailed and thorough discussion of this particular transformation, we refer the Reader to the paper \cite{SJFletcher_2003}.

Having expressed the Kerr metric in GBS form, we are now prepared to examine its asymptotic behavior in this coordinate system. Since we are interested only in the asymptotic regime, equation \ref{alpha} can be expressed in the large $r$ limit, with the substitution $x=1/y$, as
\begin{equation}
    \label{approx alpha}
    \begin{split}
        \alpha&\approx \int a~dy\left[1-\frac{1}{2} \left( a^2y^2+2a^2M_Ky^3  \right)  \right]\\
        &=a\left[ \frac{1}{x}-\frac{1}{2}\left( \frac{a^2}{3 x^3}+\frac{a^2M_K}{2x^4}    \right)   \right]^\infty _{\tilde{r}}\\
        &\approx -\frac{a}{\tilde{r}}+\frac{a^3}{6\tilde{r}^3}+\mathcal{O}(r^{-5}),
    \end{split}
\end{equation}
giving in the asymptotic form
\begin{equation}
    \tanh\alpha\approx \alpha-\frac{\alpha^3}{3}+\mathcal{O}(\alpha^5)= -\frac{a}{\tilde{r}}+\frac{a^3}{2\tilde{r}^3}+\mathcal{O}(r^{-5}).
\end{equation}

Thus, the metric components in equation \ref{GBS_kerr}, for asymptotically large $r$ values, can be expressed as
\begin{equation}
    \label{metric in r1}
    \begin{split}
        g_{ur}&=-\left[ 1-\frac{a^2}{\tilde{r}}\left( \frac{1}{2}-\cos^2\theta \right)   \right]+\mathcal{O}(\tilde{r}^{-3}),\\
        g_{u\theta}&=-\left[a \cos\theta+2a\cos\theta\left[ a\sin\theta-M_K \right]\frac{1}{\tilde{r}}\right]+\mathcal{O}(\tilde{r}^{-2}),\\
        g_{u\phi}&=-2aM_K\sin^2\theta\frac{1}{\tilde{r}}+\mathcal{O}(\tilde{r}^{-2}),\\
        g_{\theta\theta}&=\tilde{r}^2+2a\sin\theta\tilde{r}+a^2(3\sin^2\theta-1)+\mathcal{O}(\tilde{r}^{-1}),\\
        g_{\phi\phi}&=\tilde{r}^2\sin\theta^2-2a\tilde{r}\sin\theta\cos^2\theta+a^2(1-3\sin^2\theta\cos^2\theta)+\mathcal{O}(\tilde{r}^{-1}),\\
         g_{\theta\phi}&=\mathcal{O}(\tilde{r}^{-1}).
    \end{split}
\end{equation}

Finally, the BMS gauge is obtained by defining a new radial coordinate $r$ satisfying $\det[g_{AB}]=r^4\sin^2\theta$, so that
\begin{equation}
    \label{r}
    \tilde{r}=r+\frac{a}{2}\frac{\cos(2\theta)}{\sin\theta}+\frac{a^2}{8}\left[ 4\cos(2\theta)+\frac{1}{\sin^2\theta}  \right]\frac{1}{r}+\mathcal{O}(r^{-2}),
\end{equation}
yielding the metric components, in the large $r$ limit, as
\begin{equation}
    \label{metric in r}
    \begin{split}
     &g_{ur}=-\left[ 1-\frac{a^2}{r}\left( \frac{1}{2}-\cos^2\theta \right)   \right]+\mathcal{O}(r^{-3})\\
        &g_{u\theta}=\frac{a\cos\theta}{2\sin^2\theta}+\frac{a\cos\theta}{4}\left(8M_K+\frac{a}{\sin^3\theta}\right)\frac{1}{r}+\mathcal{O}(r^{-2})\\
        &g_{u\phi}=-2aM_K\sin^2\theta\frac{1}{r}+\mathcal{O}(r^{-2})\\
        &g_{\theta\theta}=r^2+\frac{ar}{\sin\theta}+\frac{a^2}{2\sin^2\theta}+\mathcal{O}(r^{-1})\\
        &g_{\phi\phi}=r^2\sin\theta^2-ar\sin\theta+\frac{a^2}{2}
        +\mathcal{O}(r^{-1})\\
         &g_{\theta\phi}=\mathcal{O}(r^{-1}).   
    \end{split}
\end{equation}

Thus, the Kerr metric, in the asymptotic limit, in this new coordinate, is given by \cite{Barnich:2011mi}

\begin{equation}
    \label{kerr_final}
    \begin{split}
        d\tilde{s}^2_{\text{KGBS}}\approx&-du^2-2dudr+r^2\Big(d\theta^2+\sin^2\theta d\phi^2\Big)+\frac{2M_K}{r}du^2+r\left(\frac{a}{\sin\theta}\right)d\theta^2+r(-a\sin\theta)d\phi^2\\
        &+2\left[\frac{a\cos\theta}{2\sin^2\theta} +\frac{2}{3r}\left\{\frac{3a\cos\theta}{8}\left(8M_K+\frac{a}{\sin^2\theta}\right) \right\}\right]dud\theta+2\left[ \frac{2}{3r}\left(  -3aM_K\sin^2\theta\right)  \right]dud\phi\\
        &=-du^2-2dudr+r^2(d\theta^2+\sin^2\theta d\phi^2)+\frac{2M_K}{r}du^2+rC^K_{\theta\theta}d\theta^2+rC^K_{\phi\phi}d\phi^2\\
        &+2g_{u\theta}dud\theta+2g_{u\phi}dud\phi.
    \end{split}
\end{equation}

In summary, expressing the Kerr solution in Bondi gauge provides a transparent view of its asymptotic structure, making explicit the angular dependence and falloff properties of the metric components near null infinity. This formulation not only facilitates the identification of asymptotic charges and symmetries but also serves as a natural starting point for obtaining an asymptotic Kerr like solution from a different geometry. With the Kerr metric in this gauge established, we are now positioned to investigate the action of supertranslations on the non-rotating background of the Schwarzschild spacetime, and examine how these transformations can yield a Kerr like asymptotic configuration at null infinity.

\section{Kerr-like Geometry from Schwarzschild via Supertranslations} \label{sec-sch}

In this section, we construct explicit supertranslation transformations that map the Schwarzschild spacetime to the asymptotically Kerr-like geometry. Utilizing the infinite-dimensional feature of the BMS group, we demonstrate how successive gauge transformations, specifically supertranslations, can interpolate between these two distinct vacuum solutions of the Einstein field equations. By carefully analyzing the required gauge conditions and large-$r$ falloff conditions, we identify the class of supertranslation functions that encode the rotational features characteristic of the Kerr metric in the asymptotic regime. This construction provides a concrete realization of how asymptotic symmetries enrich the solution space of general relativity, linking non-rotating and rotating black holes through physically meaningful soft degrees of freedom.

In order to match the Bondi-Sachs form of the Kerr metric \ref{kerr_final}  with the corresponding Schwarzschild geometry in the large-$r$ limit, we now turn to the Schwarzschild metric expressed in $(u,r,\theta,\phi)$ coordinates,
\begin{equation}
    \label{sch metric}
    ds^2_{\rm{Sch}}=-du^2-2dudr+r^2(d\theta^2+\sin^2\theta d\phi^2)+\frac{2M_s}{r}du^2.
\end{equation}

As a first step towards the matching of the two geometries, we confine our attention to a special class of diffeomorphisms that maintain the asymptotic structure of the Schwarzschild metric in equation \ref{sch metric}, and satisfies the following set of conditions:
\begin{equation}
    \label{1Lie_conditions}
    \begin{split}
        \mathcal{L}_{\zeta} g_{uu}= &~ 0,\\
        \mathcal{L}_\zeta g_{rr}= &~0,\\
        \mathcal{L}_{\zeta} g_{rA}= &~0,\\
        \mathcal{L}_{\zeta} g_{ur}= &~0,\\
    \end{split}
\end{equation}
with the large-$r$ falloffs for the vector $\zeta$ as 
\begin{equation}
    \label{zeta fall offs}
    \zeta^u \sim \mathcal{O}(1), ~~~~\zeta^r,~\zeta^A\sim \mathcal{O}(r^{-1}),
\end{equation}
where $A=\theta,\phi$ indicates the spherical coordinates.

The vector field satisfying these conditions, as shown in Appendix \ref{zeta_Lie}, has the form
\begin{equation}
    \label{zeta}
    \zeta=f\partial_u+(\frac{2M_s}{r}-1)f\partial_r-\frac{1}{r}\left[\partial_\theta f ~\partial_\theta +\frac{\partial_\phi f}{\sin^2\theta}\partial_\phi \right] ,
\end{equation}
where $f(\theta,\phi)$ is an arbitrary function of $(\theta,\phi)$. It turns out that we can further choose this function $f(\theta,\phi)$, without loss of generality, to be a Killing vector on the 2-sphere, satisfying 
\begin{equation}
    \label{zeta_killing}
    \mathcal{L}_{\zeta}g_{\theta\phi}=0~,
\end{equation}
leading to 
\begin{equation}
f(\theta,\phi)=h(\theta,\phi) \,\sin\theta ,
\end{equation}
where $h(\theta,\phi)$ is a new unknown function of $(\theta,\phi)$.

Thus, the metric for the Schwarzschild black hole, upon supertranslating by the diffeomorphic vector field $\zeta$, parametrized by the unknown function $f(\theta,\phi)$, is given by

\begin{equation}
        \label{zeta sch}
    \begin{split}
        ds^2_f&=\left( g_{\mu\nu} +\mathcal{L}_f g_{\mu\nu} \right)dx^\mu dx^\nu \\
        &=-du^2-2dudr+r^2(d\theta^2+\sin^2\theta d\phi^2)
        +\frac{2M_s}{r}du^2+rC^s_{\theta\theta}d\theta^2+rC^s_{\phi\phi}d\phi^2,
    \end{split}
\end{equation}

with
\begin{equation}
\label{C_S}
    \begin{split}
        C^s_{\theta\theta}&=-2(f+\partial_\theta^2f),\\
        C^s_{\phi\phi}&=-2\left( f\sin^2\theta+\sin\theta\cos\theta\partial_\theta f+\partial_\phi^2f \right).
    \end{split}
\end{equation}
Interestingly, the imprint of supertranslation is encoded in the components $C^s_{\theta\theta}$ and $C^s_{\phi\phi}$, which are directly associated with gravitational wave degrees of freedom.

As a final step, we now consider the second class of diffeomorphisms that preserve both the standard Bondi gauge conditions and the large-$r$ falloff behavior of the modified metric given in Eq.~\eqref{zeta sch}, namely,
\begin{equation}
    \label{eta conditions}
    \begin{split}
        \mathcal{L}_\eta g_{rr}&=~0,\\
        \mathcal{L}_{\eta} g_{rA}&=~0,\\
        \gamma^{AB}\mathcal{L}_{\eta}g_{AB}&=~0,\\
        \mathcal{L}_{\eta} g_{ur}&=~0,\\
    \end{split}
\end{equation}
and the falloff conditions, 
\begin{equation}
    \label{eta falloff}
    \eta^u, \eta^r\sim \mathcal{O}(1);~~~~~\eta^\theta,~\eta^\phi\sim \mathcal{O}(r^{-1}).
\end{equation}

As shown in Appendix \ref{eta app}, the vector field $\eta$ satisfying the above conditions is given by
\begin{equation}
\label{eta_final sln}
    \eta=F\partial_u+\frac{1}{2}D^2F \partial_r-\frac{1}{r}\left[\partial_\theta F \partial_\theta+\frac{\partial_\phi F}{\sin^2\theta} \partial_\phi \right],
\end{equation}
with 
\begin{equation}
    \label{laplacian}
    D^2\mathcal{F}(\theta,\phi)=\frac{1}{\sin\theta}\partial_\theta\left( \sin\theta\partial_\theta\mathcal{F} \right)+\frac{1}{\sin^2\theta}\partial_\phi^2\mathcal{F}
\end{equation}
representing the Laplacian of any function $\mathcal{F}(\theta,\phi)$ on the $2$-sphere.

Thus, the supertranslated Schwarzschild metric \ref{zeta sch}, upon supertranslating again by the vector field $\eta$, assumes the form 

\begin{equation}
    \label{final sch}
    \begin{split}
        ds^2_{f+F}&=-du^2-2dudr+r^2(d\theta^2+\sin^2\theta d\phi^2)+\frac{2M_s}{r}du^2+r\Big[C^S_{\theta\theta}+D^2F -2\partial_\theta^2F+\mathcal{O}(r^{-1})  \Big]d\theta^2\\
        &+r\Bigg[C^S_{\phi\phi}+\sin^2\theta D^2F-2\sin\theta\cos\theta\partial_\theta F -2\partial_\phi^2F+\mathcal{O}(r^{-1}) \Bigg]d\phi^2\\&+2\partial_\theta\Big[ \left( \frac{2M_s}{r}-1 \right)F-\frac{1}{2}D^2F  \Big]dud\theta+2\partial_\phi\Big[ \left( \frac{2M_s}{r}-1 \right)F-\frac{1}{2}D^2F  \Big]dud\phi.
    \end{split}
\end{equation}

It is important to note that $ds^2_{f+F}$ represents the Schwarzschild metric having supertranslated first with respect to the vector field $\zeta$, parametrized by $f(\theta,\phi)$, and subsequently with respect to the vector field $\eta$, parametrized by $F(\theta,\phi)$.

Thus, our construction demonstrates that by applying two successive supertranslations with appropriately chosen guage conditions and falloff conditions at large $r$, the asymptotic geometry of the Schwarzschild spacetime can be deformed into a Kerr-like configuration. This result exemplifies the power of the BMS symmetry group in relating seemingly distinct gravitational vacua and highlights the role of supertranslation hair in encoding black hole geometries. These findings provide new insights into the structure of the gravitational solution space and set the stage for further investigations into how asymptotic symmetries may unify our understanding in relating different black hole spacetimes.

\section{Matching of Supertranslated Schwarzschild and Kerr Asymptotics} \label{sec-match}

In this section, we present a detailed mathematical construction that establishes the explicit correspondence between the Schwarzschild metric subjected to successive supertranslations and the asymptotic structure characteristic of the Kerr geometry. By carefully analyzing the large-$r$ expansions with appropriate gauge conditions, we demonstrate how the two sets of metric components align, thereby validating the role of BMS supertranslations as transformations that interpolate between non-rotating and rotating black hole spacetimes. This explicit matching highlights the deep interplay between asymptotic symmetries and classical gravitational solutions, providing a concrete realization of the conceptual framework introduced in the previous sections.

Through asymptotic matching of these two black hole metrics, namely equations \ref{kerr_final} and \ref{final sch}, we find that the two supertranslation functions $f(\theta,\phi)$ and $F(\theta,\phi)$ must satisfy the following equalities:
\begin{equation}
\label{1st compared}
C^S_{\theta\theta}+D^2F-2\partial_\theta^2F=C^K_{\theta\theta},
\end{equation}
\begin{equation}
    \label{2nd compared}
    C^S_{\phi\phi}+\sin^2\theta D^2F-2\sin\theta\cos\theta\partial_\theta F-2\partial^2_\phi F=C^K_{\phi\phi},
\end{equation}
\begin{equation}
\label{3rd compared}
    \partial_\theta\left[ F+\frac{1}{2}D^2F  \right]=-g^K_{u\theta},
\end{equation}
and
\begin{equation}
    \label{4th compared}
    \partial_\phi\left[ F+\frac{1}{2}D^2F  \right]=-g^K_{u\phi}.
\end{equation}

Noting from \ref{kerr_final} that $g^K_{u\phi}=0$ and $g^K_{u\theta}=\frac{a\cos\theta}{2\sin^2\theta}$, Equations \ref{3rd compared} and \ref{4th compared} together imply
\begin{equation}
    \label{F equation}
    D^2F+2F=\frac{a}{\sin\theta}+C
\end{equation}
where $C$ is the integration constant.

Furthermore, Equations \ref{1st compared} and \ref{2nd compared}, upon extracting their right hand sides from equation \ref{kerr_final}, and using the identifications given by Equations \ref{C_S}, lead to
\begin{equation}
    \label{f1 equation}
  -2(f+\partial_\theta^2f)+\left[\cot\theta\partial_\theta F-\partial_\theta^2F+\frac{\partial_\phi^2 F}{\sin^2\theta} \right]=\frac{a}{\sin\theta}
\end{equation}
and

\begin{equation}
    \label{f2 equation}
   \left[\cot\theta\partial_\theta F-\partial_\theta^2F+\frac{\partial_\phi^2 F}{\sin^2\theta} \right]-\frac{a}{\sin\theta}=-2\left(f+\cot\theta\partial_\theta f+\frac{\partial_\phi^2 f}{\sin^2\theta}   \right).
\end{equation}

Equations \ref{f1 equation} and \ref{f2 equation} together imply
\begin{equation}
    \label{f equation}
    D^2f+2f=0,
\end{equation}
whose solution is readily obtained as
\begin{equation}
    \label{f solution}
    f(\theta,\phi)=\sum_{m=-l}^{m=l}A_m P_l^m(\cos\theta) e^{im\phi}, ~~\text{with}~~l=1,
\end{equation}
where $P_l^m(\cos\theta)$ is the associated Legendre function, given by the Rodrigues formula
\begin{equation}
    \label{rodrig}
    P_l^m(x)=(-1)^m(1-x^2)^{m/2}\frac{d^m}{dx^m}P_l(x),
\end{equation}
with $P_l(x)$ the Legendre polynomial of degree $l$.

Turning to Equation \ref{F equation}, its complementary function is given by
\begin{equation}
    \label{F_CF solution}
    F_{CF}(\theta,\phi)=\sum_{m=-l}^{m=l}B_m P_l^m(\cos\theta) e^{im\phi}, ~~\text{with}~~l=1.
\end{equation}

To obtain the particular integral of equation \ref{F equation}, we series expand both sides in terms of the spherical harmonics. Since the right-hand side of equation \ref{F equation} is independent of $\phi$, we can expand both sides in terms of Legendre polynomials, with the eigen value equation $D^2 P_n(\cos\theta)=-n(n+1)P_n(\cos\theta)$.
Thus, assuming $F_{PI}(\theta,\phi)=F_{PI}(\theta)=\sum_{n=0}^\infty a_n P_n(\cos\theta)$, the left-hand side of equation  \ref{F equation} reads
\begin{equation}
    \label{lhs}
    D^2F+2F=\sum_{n=0}^\infty a_n\left[2-n(n+1) \right]P_n(\cos\theta),
\end{equation}
whereas, with $C=C P_0(\cos\theta)$, the right-hand side can be expressed as
\begin{equation}
    \label{rhs}
    C+\frac{a}{\sin\theta}=C P_0(\cos\theta) +\frac{a\pi}{2}\sum_{k=0}^\infty\frac{(2k-1)!!}{(2k)!!}P_{2k}(\cos\theta),
\end{equation}
where $(2k-1)!!=(2k-1)(2k-3)\dots 3\cdot1$ and $(2k)!!=2k(2k-2)\dots 4\cdot2$.

Expressing the right-hand side as $\sum_{n=0}^\infty b_n P_n(\cos\theta)$, we have
\begin{equation}
b_n =
\begin{cases}
\displaystyle C+\frac{a\pi}{2}~; ~~n=0 \\[1.2em]
\displaystyle \frac{a\pi}{2}\sum_{k=0}^\infty \frac{(2k-1)!!}{(2k)!!}~; ~~n=2k\geq2\\
\displaystyle 0~~; n\in\text{odd}
\end{cases}
\end{equation}

Thus, we have
\begin{equation}
    \label{lhs=rhs}
    \sum_{n=0}^\infty a_n \left[2-n(n+1) \right]P_n(\cos\theta)=\sum_{n=0}^\infty b_n P_n(\cos\theta)
\end{equation}
giving
\begin{equation}
    a_n =
\begin{cases}
\displaystyle \frac{2C+a\pi}{4}~; ~~n=0 \\[1.2em]
\displaystyle \frac{a\pi}{2} \frac{(2k-1)!!}{(2k)!![k(2k+1)-1]}~; ~~n=2k\geq2\\
\displaystyle 0~~; n=\text{Odd}
\end{cases}
\end{equation}
Therefore, the complete solution is given by

\begin{equation}
    \label{F solution}
    F(\theta,\phi)=\sum_{m=-1}^{m=1}B_m P_1^m(\cos\theta) e^{im\phi}+\frac{2C+a\pi}{4}-\sum_{k=1}^\infty \frac{a\pi}{4} \frac{(2k-1)!!}{(2k)!![k(2k+1)-1]} P_{2k}(\cos\theta). 
\end{equation}

The first supertranslation function $f(\theta,\phi)$, corresponding to the vector field $\zeta$ is obtained as the $l = 1$ spherical harmonics, given by equation \ref{f solution}. On the other hand, the complete solution for $F(\theta,\phi)$, given in equation \ref{F solution}, contains an infinite series of Legendre polynomials of even order $l \geq 2$, in addition to the $l = 1$ spherical harmonics, and a constant term.

The two families of supertranslations highlight distinct physical features. The first supertranslation function $f(\theta,\phi)$, with $l = 1$, corresponds to a center of mass displacement or boost in the Bondi frame, which does not introduce any new hair, is associated with the displacement memory and BMS translations  \cite{Strominger:2014pwa,Flanagan:2014kfa}.  The second supertranslation function $F(\theta,\phi)$, characterized by even ordered Legendre polynomials, is related to deeper intrinsic properties of the Kerr geometry. The higher order modes contained in the second supertranslation  $F(\theta,\phi)$ include even $l \geq 2$ components, which directly relates to the mass type multipole structure of Kerr black holes.

\section{Discussion and Conclusion} \label{disc} 

With the development of Bondi-van der Burg-Metzner-Sachs (BMS) symmetry algebra and with its potential to address some of the fundamental questions of general relativity and black hole physics, it becomes inevitable to check if the different geometries of spacetime, particularly different solutions of the Einstein field equation, are fundamentally different or are connected to each other in the asymptotic limit through some gauge symmetries. 

To address this question, we investigated  in this work the connection between Schwarzschild and Kerr black holes within the framework of BMS symmetries. By implementing two successive supertranslations on the Schwarzschild spacetime, we showed that its asymptotic structure can be matched with that of the Kerr geometry. The analysis revealed two distinct families of supertranslation functions, each playing a different physical role in the transformation.

As discussed earlier, the first supertranslation function $f(\theta,\phi)$, determined by $l=1$ spherical harmonics, corresponds to a global translation in the Bondi frame without introducing any new hair. However, the second supertranslation function $F(\theta,\phi)$ includes all even $l \geq 2$ components of Legendre polynomials, which captures the mass type multipole structure of the Kerr black hole.

Together, these two kinds of supertranslations provide a novel insight about the nature of different black hole geometries offering a unified description of soft hair across both rotating and nonrotating black holes. This correspondence shows that supertranslations contain physically meaningful soft degrees of freedom and provides a natural way to describe soft hair on both rotating and non-rotating black holes.

In Thorne or Geroch-Hansen formalism, the multipole expansion of Kerr spacetime satisfies the relation $M_l + iS_l = M (ia)^l$, which implies that only even mass multipoles $(M_{2k})$ and odd current multipoles $(S_{2k+1})$ survive \cite{Thorne:1980ru, Hansen:1974zz}. This selection rule is a direct result of Kerr’s equatorial reflection symmetry, forcing all odd mass multipoles and even current multipoles to vanish. The emergence of even-order Legendre polynomials in $F(\theta,\phi)$ thus mirrors the intrinsic multipolar structure of the Kerr geometry, suggesting that this supertranslation encodes genuine soft degrees of freedom associated with rotation. Therefore, these modes provide a symmetric and controlled way to encode the soft degrees of freedom in the Kerr geometry, in agreement with their interpretation of soft hair as carriers of BMS charges \cite{Hawking:2016sgy,Donnay_2018}.

The successful analytical matching between the Schwarzschild metric subjected to successive supertranslations and the asymptotic form of the Kerr geometry provides compelling evidence that BMS supertranslations act as a bridge connecting these distinct geometries. This correspondence reinforces the notion that the space of asymptotically flat solutions to Einstein’s equations is intimately interconnected via infinite-dimensional symmetry transformations. Moreover, it underscores the physical relevance of supertranslation hair as carriers of soft gravitational degrees of freedom, offering a promising avenue for a unified description of black hole microstates and their associated multipolar structure.

\section{Acknowledgments}
Nihar Ranjan Ghosh is supported through a Research Fellowship from the Ministry of Human Resource Development (MHRD), Government of India.

\section*{Appendixes}

\appendix
\renewcommand{\theequation}{\Alph{section}\arabic{equation}}

\section{\label{zeta_Lie}}

In this section we shorly review and derive the Lie derivatives of the metric $g_{\mu\nu}$, defined with respect to a vector $\xi$ as $\mathcal{L}_{\xi}g_{\mu\nu}=\xi^\alpha\partial_\alpha g_{\mu\nu}+g_{\nu\alpha}\partial_\mu\xi^\alpha+g_{\mu\alpha}\partial_\nu\xi^\alpha$. Thus, the condition $\mathcal{L}_{\zeta}g_{rr}=0$ in equation \ref{1Lie_conditions} implies $\partial_r\zeta^u=0$, giving
\setcounter{equation}{0}
\begin{equation}
    \begin{split}
    \label{zeta_u}
         &\zeta^u\equiv f(u,\theta,\phi).
    \end{split}
\end{equation}
Similarly, the condition $\mathcal{L}_{\zeta}g_{uu}=0$ in \ref{1Lie_conditions} implies $\left(\frac{2M_s}{r}-1\right)\partial_u\zeta^u-\partial_u\zeta^r+\mathcal{O}(r^{-2})=0$, providing 
\begin{equation}
    \begin{split}
    \label{zeta_r}
     \zeta^r=\left(\frac{2M_s}{r}-1\right) f~,
    \end{split}
\end{equation}
and the third condition $\mathcal{L}_{\zeta}g_{rA}=0$ leads to $-\partial_A\zeta^u+r^2\gamma_{AA}\partial_r\zeta^A=0$, so that   
\begin{equation}
    \begin{split}
    \zeta^A=-\frac{1}{r}\frac{1}{\gamma_{AA}}\partial_Af,
    \end{split}
\end{equation} 
with $\gamma_{AA}$ the $(AA)^{\text{th}}$ component of the matrix $\gamma_{AB}$, given by 
\begin{equation}
\label{gamma}
\gamma_{AB} =
\begin{pmatrix}
1 &0 \\
0 & \sin^2\theta
\end{pmatrix},
\end{equation}
implying
\begin{equation}
    \label{zeta_A}
\zeta^A =
\begin{cases}
\displaystyle \zeta^\theta=-\frac{1}{r}\partial_\theta f \\[1.2em]
\displaystyle \zeta^\phi=-\frac{1}{r\sin^2\theta}\partial_\phi f.
\end{cases}
\end{equation}
Lastly, the condition $\mathcal{L}_{\zeta}g_{ur}=0$ gives $\frac{2M_s}{r^2}f-\partial_u f=0$, which, upon using the falloff condition given by \ref{zeta fall offs}, leads to
\begin{equation}
    \begin{split}
    \label{zeta_last}
        f \equiv f(\theta,\phi) .
    \end{split}
\end{equation}

Thus equations \ref{zeta_u}, \ref{zeta_r}, \ref{zeta_A} and \ref{zeta_last} define the vector field $\zeta$ as
\begin{equation}
    \zeta=f\partial_u+(\frac{2M_s}{r}-1)f\partial_r-\frac{1}{r}\left[\partial_\theta f ~\partial_\theta +\frac{\partial_\phi f}{\sin^2\theta}\partial_\phi \right] .
\end{equation}

In order to obtain the supertransleted Schwarzschild metric with respect to the above-obtained vector field $\zeta$, we obtain the Lie derivatives of the $\theta\theta$ component as 
\begin{equation}
        \mathcal{L}_{f}g_{\theta\theta}=\zeta^r\partial_r(r^2)+2g_{\theta\theta}\partial_{\theta}\zeta^\theta=(4M_s-2r)f-2r\partial_\theta^2f,
\end{equation}
and similarly, of the $\phi\phi$ component as 
\begin{equation}
    \begin{split}
        \mathcal{L}_fg_{\phi\phi}&=\zeta^r\partial_r(r^2)\sin^2\theta+r^2\zeta^\theta\partial_\theta(\sin^2\theta)+2r^2\sin^2\theta\partial_\phi\zeta^\phi\\
        &=(4M_s-2r)f\sin^2\theta-2r\sin\theta\cos\theta\partial_\theta f-2r\partial^2_\phi f.
    \end{split}
\end{equation}

The above results are used to obtain the supertranslated Schwarzschild metric with respect to the vector $\zeta$, given by equation \ref{zeta sch}.

\section{\label{eta app}}
\setcounter{equation}{0}
The action of the first vector field $\zeta$ changes the Schwarzschild metric to \ref{zeta sch}, where only the spherical part is modified, keeping the rest unchanged. Consequently, we introduced the second diffeomorphic vector $\eta$ that changes the $rr$ and $rA$ metric components preserving the Bondi gauge.

Following similar arguments as before, the first condition in \ref{eta conditions} gives 
\begin{equation}
      \label{eta_u}
      \eta^u\equiv F(u,\theta,\phi).
    \end{equation}

Similarly, the second and fourth conditions in \ref{eta conditions} give $-\partial_A\eta^u+r^2\left( \gamma_{AA}+\frac{C^s_{AA}}{r}  \right)\partial_r\eta^A=0$, and $-\partial_u\eta^u=0$, resulting in 
\begin{equation}
\label{eta_A}
    \eta^A=\frac{-1}{r\gamma_{AA}}\partial_AF
\end{equation}
and 
\begin{equation}
\label{eta_last}
\eta^u\equiv F(\theta,\phi),
\end{equation}
where $\gamma_{AA}$ is the $(AA)^\text{th}$ component of the matrix $\gamma_{AB}$, given in \ref{gamma}.

Finally, the third condition in \ref{eta conditions}, requires its calculation before taking the trace with respect to the original $2$-sphere metric:
\begin{equation}
    \begin{split}
        \mathcal{L}_{\eta}g_{AB}&=\eta^\alpha\partial_\alpha g_{AB}+g_{A\alpha}\partial_B\eta^\alpha+g_{\alpha B}\partial_A\eta^\alpha\\
        &=\eta^\alpha\partial_\alpha\left(r^2\gamma_{AB}+rC^s_{AB}  \right)+\left(r^2\gamma_{AC}+rC^s_{AC} \right)\partial_B\eta^C+\left(r^2\gamma_{BC}+rC^s_{BC} \right)\partial_A\eta^C\\
        &=\left(2r\gamma_{AB}+C^s_{AB} \right)\eta^r+r^2\eta^c\partial_C \gamma_{AB}+r\eta^C\partial_CC^s_{AB}+r^2\left(\gamma_{AC}\partial_B\eta^C+\gamma_{CB}\partial_A\eta^C\right)\\&~~~+r\left(C^s_{AC}\partial_B\eta^C+C^s_{CB}\partial_A\eta^C\right)
    \end{split}
\end{equation}

Thus, taking the trace $\gamma^{AB}\mathcal{L}_{\eta}g_{AB}=0$, the above equation gives
\begin{equation}
\label{eta_r1}
	(4r+C^A_A)\eta^r+r^2\left(\gamma^{AB}\eta^C\partial_C\gamma_{AB}+2\partial_A\eta^A\right)+r\left( 2\gamma^{AB}C_{AC}\partial_B\eta^C+\gamma^{AB}		\eta^C\partial_CC^S_{AB}  \right)=0,
\end{equation}
that implies
\begin{equation}
\label{eta_r2}
    \begin{split}
        &-(4r+C^A_A)\eta^r=-r\left(2\cot\theta\partial_\theta F+2\partial^2_\theta F+\frac{2}{\sin^2\theta}\partial^2_\phi F   \right)\\
        &+r\Bigg[2\left\{  \gamma^{\theta\theta} C^S_{\theta\theta}\partial_\theta\eta^\theta+\gamma^{\phi\phi}C^S_{\phi\phi}\partial_\phi\eta^\phi    \right\}+\gamma^{\theta\theta}\left( \eta^\theta\partial_\theta C^S_{\theta\theta}+\eta^\phi \partial_\phi C^S_{\theta\theta} \right)+  \gamma^{\phi\phi}\left( \eta^\theta\partial_\theta C^S_{\phi\phi}+\eta^\phi \partial_\phi C^S_{\phi\phi} \right)                \Bigg]\\
        &=-2r D^2F-\Bigg[2\left( C^S_{\theta\theta}\partial^2_\theta F+\frac{C^S_{\phi\phi}}{\sin^4\theta}\partial^2_\phi F \right)+\left(\partial_\theta F\partial_\theta C^S_{\theta\theta} +\frac{1}{\sin^2\theta}\partial_\phi C^S_{\theta\theta}  \right)\\&+\frac{1}{\sin^2\theta}\left( \frac{1}{\sin\theta} \partial_\theta C^S_{\phi\phi} \frac{1}{\sin^2\theta}\partial_\phi C^S_{\phi\phi}\right)      \Bigg],
    \end{split}
\end{equation}
which simplifies to $ (4r+C^A_A)\eta^r=2rD^2 F+G(u,\theta,\phi)$, giving the solution for $\eta^r$ as:
\begin{equation}
\label{eta_r}
\eta^r \Big|_{r\to\infty}=\frac{1}{2}D^2 F.
\end{equation}

Consequently, equations \ref{eta_u}, \ref{eta_A}, \ref{eta_last}  and \ref{eta_r} give the vector field as 
\begin{equation}
\eta=F\partial_u+\frac{1}{2}D^2F \partial_r-\frac{1}{r}\left[\partial_\theta F \partial_\theta+\frac{\partial_\phi F}{\sin^2\theta} \partial_\phi \right],
\end{equation}
which is also given in equation \ref{eta_final sln}.

Interestingly, the vector field $\eta$ has the same asymptotic structure as one would obtain upon starting from the Schwarzschild metric with the same gauge and large $r$ falloff conditions for $\eta$.  

Now, the action of the diffeomorphic vector field $\eta$ on the $\theta\theta$ component of the metric can be obtained as
\begin{equation}
    \begin{split}
        \mathcal{L}_{\eta}g_{\theta\theta}&=\eta^\alpha\partial_\alpha(r^2+r C^S_{\theta\theta})+2g_{\theta\alpha}\partial_\theta\eta^\alpha\\
        &=(2r+C^S_{\theta\theta})\eta^r+r(\eta^\theta\partial_\theta C^S_{\theta\theta}+\eta^\phi\partial_\phi C^S_{\theta\theta})+2g_{\theta\theta}\partial_\theta\eta^\theta\\
        &=(2r+C^S_{\theta\theta})\frac{1}{2}D^2F-(\partial_\theta F\partial_\theta C^S_{\theta\theta}+\frac{1}{\sin^2\theta}\partial_\phi F\partial_\phi C^S_{\theta\theta})-2(r+C^S_{\theta\theta})\partial_\theta^2 F\\
        &=r(D^2F-2\partial_\theta^2 F)+\left[\frac{1}{2}C^S_{\theta\theta}D^2 F-\partial_\theta F\partial_\theta C^S_{\theta\theta}-\frac{1}{\sin^2\theta}\partial_\phi F \partial_\phi C^S_{\theta\theta}-2C^S_{\theta\theta}\partial_\theta^2 F\right],
    \end{split}
\end{equation}
and, similarly, for the $\phi\phi$ component as
\begin{equation}
    \begin{split}
        &\mathcal{L}_\eta g_{\phi\phi}=\eta^\alpha\partial_\alpha(r^2\sin^2\theta+rC^S_{\phi\phi})+2g_{\phi\alpha}\partial_\phi\eta^\alpha\\
        &=(2r\sin^2\theta+C^S_{\phi\phi})\eta^r+2r^2\sin\theta\cos\theta\eta^\theta+r(\eta^\theta\partial_\theta C^S_{\phi\phi}+\eta^\phi\partial_\phi C^S_{\phi\phi})-\frac{2}{\sin^2\theta}(r\sin^2\theta+C^S_{\phi\phi})\partial_\phi^2 F\\
        &=(2r\sin^2\theta+C^S_{\phi\phi})\frac{1}{2}D^2F-2r\sin\theta\cos\theta\partial_\theta F-\Big(\partial_\theta F\partial_\theta C^S_{\phi\phi}+\frac{1}{\sin^2\theta}\partial_\phi C^S_{\phi\phi}\Big)\\
        & ~~~  -\frac{2}{\sin^2\theta}(r\sin^2\theta+C^S_{\phi\phi})\partial_\phi^2 F\\
        &=r\Big[\sin^2\theta D^2F-2\sin\theta\cos\theta \partial_\theta F-2\partial_\phi^2F\Big]+\Bigg[\frac{1}{2}C^S_{\phi\phi}D^2F-\partial_\theta F\partial_\theta C^S_{\phi\phi}-\frac{1}{\sin^2\theta}\partial_\phi F\partial_\phi C^S_{\phi\phi}\\
        & ~~~ -\frac{2}{\sin^2\theta}C^S_{\phi\phi}\partial_\phi^2 F\Bigg].
    \end{split}
\end{equation}

Lastly, the remaining components modify as: 
\begin{equation}
      \mathcal{L}_\eta g_{Au}=g_{uu}\partial_A\eta^u+g_{ur}\partial_A\eta^r+g_{AA}\partial_u\eta^A
      =\partial_A\Big[ \left(\frac{2M_s}{r}-1\right)F-\frac{1}{2}D^2F  \Big].
\end{equation}

The above results are used to obtain the supertranslated Schwarzschild metric with respect to the vector $\eta$, given by equation \ref{final sch}.


\begin{thebibliography}{10}

\bibitem{bondi1962gravitational}
Hermann Bondi, M~Gr~J Van~der Burg, and AWK Metzner.
\newblock Gravitational waves in general relativity, vii. waves from
  axi-symmetric isolated system.
\newblock {\em Proceedings of the Royal Society of London. Series A.
  Mathematical and Physical Sciences}, 269(1336):21--52, 1962.

\bibitem{sachs1962gravitational}
Rainer~K Sachs.
\newblock Gravitational waves in general relativity viii. waves in
  asymptotically flat space-time.
\newblock {\em Proceedings of the Royal Society of London. Series A.
  Mathematical and Physical Sciences}, 270(1340):103--126, 1962.

\bibitem{strominger2018lecturesinfraredstructuregravity}
Andrew Strominger.
\newblock Lectures on the infrared structure of gravity and gauge theory, 2018.
\newblock URL: \url{https://arxiv.org/abs/1703.05448}, \href
  {http://arxiv.org/abs/1703.05448} {\path{arXiv:1703.05448}}.

\bibitem{brown1986central}
J~David Brown and Marc Henneaux.
\newblock Central charges in the canonical realization of asymptotic
  symmetries: An example from three dimensional gravity.
\newblock {\em Communications in Mathematical Physics}, 104(2):207--226, 1986.

\bibitem{CADONI1999165}
Mariano Cadoni and Salvatore Mignemi.
\newblock Asymptotic symmetries of ads2 and conformal group in d = 1.
\newblock {\em Nuclear Physics B}, 557(1):165--180, 1999.
\newblock URL:
  \url{https://www.sciencedirect.com/science/article/pii/S0550321399003983},
  \href {https://doi.org/https://doi.org/10.1016/S0550-3213(99)00398-3}
  {\path{doi:https://doi.org/10.1016/S0550-3213(99)00398-3}}.

\bibitem{hotta1998asymptoticisometrydimensionalantide}
Masahiro Hotta.
\newblock Asymptotic isometry and two dimensional anti-de sitter gravity, 1998.
\newblock URL: \url{https://arxiv.org/abs/gr-qc/9809035}, \href
  {http://arxiv.org/abs/gr-qc/9809035} {\path{arXiv:gr-qc/9809035}}.

\bibitem{henneaux1985asymptotically}
Marc Henneaux and Claudio Teitelboim.
\newblock Asymptotically anti-de sitter spaces.
\newblock {\em Communications in Mathematical Physics}, 98(3):391--424, 1985.

\bibitem{Comp_re_2016}
G.~Compère, P.~Mao, A.~Seraj, and M.~M. Sheikh-Jabbari.
\newblock Symplectic and killing symmetries of ads3 gravity: holographic vs
  boundary gravitons.
\newblock {\em Journal of High Energy Physics}, 2016(1), January 2016.
\newblock URL: \url{http://dx.doi.org/10.1007/JHEP01(2016)080}, \href
  {https://doi.org/10.1007/jhep01(2016)080}
  {\path{doi:10.1007/jhep01(2016)080}}.

\bibitem{Zeldovich:1974gvh}
Y.~B. Zel'dovich and A.~G. Polnarev.
\newblock {Radiation of gravitational waves by a cluster of superdense stars}.
\newblock {\em Sov. Astron.}, 18:17, 1974.

\bibitem{Christodoulou}
Demetrios Christodoulou.
\newblock Nonlinear nature of gravitation and gravitational-wave experiments.
\newblock {\em Phys. Rev. Lett.}, 67:1486--1489, Sep 1991.
\newblock URL: \url{https://link.aps.org/doi/10.1103/PhysRevLett.67.1486},
  \href {https://doi.org/10.1103/PhysRevLett.67.1486}
  {\path{doi:10.1103/PhysRevLett.67.1486}}.

\bibitem{Braginsky:1985vlg}
V.~B. Braginsky and L.~P. Grishchuk.
\newblock {Kinematic Resonance and Memory Effect in Free Mass Gravitational
  Antennas}.
\newblock {\em Sov. Phys. JETP}, 62:427--430, 1985.

\bibitem{Braginsky:1987kwo}
Vladimir~B. Braginsky and Kip~S. Thorne.
\newblock {Gravitational-wave bursts with memory and experimental prospects}.
\newblock {\em Nature}, 327:123--125, 1987.
\newblock \href {https://doi.org/10.1038/327123a0}
  {\path{doi:10.1038/327123a0}}.

\bibitem{PhysRevD.44.R2945}
Alan~G. Wiseman and Clifford~M. Will.
\newblock Christodoulou's nonlinear gravitational-wave memory: Evaluation in
  the quadrupole approximation.
\newblock {\em Phys. Rev. D}, 44:R2945--R2949, Nov 1991.
\newblock URL: \url{https://link.aps.org/doi/10.1103/PhysRevD.44.R2945}, \href
  {https://doi.org/10.1103/PhysRevD.44.R2945}
  {\path{doi:10.1103/PhysRevD.44.R2945}}.

\bibitem{PhysRevD.46.4304}
Luc Blanchet and Thibault Damour.
\newblock Hereditary effects in gravitational radiation.
\newblock {\em Phys. Rev. D}, 46:4304--4319, Nov 1992.
\newblock URL: \url{https://link.aps.org/doi/10.1103/PhysRevD.46.4304}, \href
  {https://doi.org/10.1103/PhysRevD.46.4304}
  {\path{doi:10.1103/PhysRevD.46.4304}}.

\bibitem{PhysRevD.45.520}
Kip~S. Thorne.
\newblock Gravitational-wave bursts with memory: The christodoulou effect.
\newblock {\em Phys. Rev. D}, 45:520--524, Jan 1992.
\newblock URL: \url{https://link.aps.org/doi/10.1103/PhysRevD.45.520}, \href
  {https://doi.org/10.1103/PhysRevD.45.520}
  {\path{doi:10.1103/PhysRevD.45.520}}.

\bibitem{Favata:2010zu}
Marc Favata.
\newblock {The gravitational-wave memory effect}.
\newblock {\em Class. Quant. Grav.}, 27:084036, 2010.
\newblock \href {http://arxiv.org/abs/1003.3486} {\path{arXiv:1003.3486}},
  \href {https://doi.org/10.1088/0264-9381/27/8/084036}
  {\path{doi:10.1088/0264-9381/27/8/084036}}.

\bibitem{Tolish:2014bka}
Alexander Tolish and Robert~M. Wald.
\newblock {Retarded Fields of Null Particles and the Memory Effect}.
\newblock {\em Phys. Rev. D}, 89(6):064008, 2014.
\newblock \href {http://arxiv.org/abs/1401.5831} {\path{arXiv:1401.5831}},
  \href {https://doi.org/10.1103/PhysRevD.89.064008}
  {\path{doi:10.1103/PhysRevD.89.064008}}.

\bibitem{Tolish:2014oda}
Alexander Tolish, Lydia Bieri, David Garfinkle, and Robert~M. Wald.
\newblock {Examination of a simple example of gravitational wave memory}.
\newblock {\em Phys. Rev. D}, 90(4):044060, 2014.
\newblock \href {http://arxiv.org/abs/1405.6396} {\path{arXiv:1405.6396}},
  \href {https://doi.org/10.1103/PhysRevD.90.044060}
  {\path{doi:10.1103/PhysRevD.90.044060}}.

\bibitem{Winicour:2014ska}
J.~Winicour.
\newblock {Global aspects of radiation memory}.
\newblock {\em Class. Quant. Grav.}, 31:205003, 2014.
\newblock \href {http://arxiv.org/abs/1407.0259} {\path{arXiv:1407.0259}},
  \href {https://doi.org/10.1088/0264-9381/31/20/205003}
  {\path{doi:10.1088/0264-9381/31/20/205003}}.

\bibitem{Strominger:2014pwa}
Andrew Strominger and Alexander Zhiboedov.
\newblock {Gravitational Memory, BMS Supertranslations and Soft Theorems}.
\newblock {\em JHEP}, 01:086, 2016.
\newblock \href {http://arxiv.org/abs/1411.5745} {\path{arXiv:1411.5745}},
  \href {https://doi.org/10.1007/JHEP01(2016)086}
  {\path{doi:10.1007/JHEP01(2016)086}}.

\bibitem{PhysRevD.92.084057}
\'Eanna~\'E. Flanagan and David~A. Nichols.
\newblock Observer dependence of angular momentum in general relativity and its
  relationship to the gravitational-wave memory effect.
\newblock {\em Phys. Rev. D}, 92:084057, Oct 2015.
\newblock URL: \url{https://link.aps.org/doi/10.1103/PhysRevD.92.084057}, \href
  {https://doi.org/10.1103/PhysRevD.92.084057}
  {\path{doi:10.1103/PhysRevD.92.084057}}.

\bibitem{Flanagan:2015pxa}
{\'E}anna~{\'E}. Flanagan and David~A. Nichols.
\newblock {Conserved charges of the extended Bondi-Metzner-Sachs algebra}.
\newblock {\em Phys. Rev. D}, 95(4):044002, 2017.
\newblock [Erratum: Phys.Rev.D 108, 069902 (2023)].
\newblock \href {http://arxiv.org/abs/1510.03386} {\path{arXiv:1510.03386}},
  \href {https://doi.org/10.1103/PhysRevD.95.044002}
  {\path{doi:10.1103/PhysRevD.95.044002}}.

\bibitem{Pasterski:2015tva}
Sabrina Pasterski, Andrew Strominger, and Alexander Zhiboedov.
\newblock {New Gravitational Memories}.
\newblock {\em JHEP}, 12:053, 2016.
\newblock \href {http://arxiv.org/abs/1502.06120} {\path{arXiv:1502.06120}},
  \href {https://doi.org/10.1007/JHEP12(2016)053}
  {\path{doi:10.1007/JHEP12(2016)053}}.

\bibitem{Pasterski:2015zua}
Sabrina Pasterski.
\newblock {Asymptotic Symmetries and Electromagnetic Memory}.
\newblock {\em JHEP}, 09:154, 2017.
\newblock \href {http://arxiv.org/abs/1505.00716} {\path{arXiv:1505.00716}},
  \href {https://doi.org/10.1007/JHEP09(2017)154}
  {\path{doi:10.1007/JHEP09(2017)154}}.

\bibitem{PhysRevD.98.064032}
David~A. Nichols.
\newblock Center-of-mass angular momentum and memory effect in asymptotically
  flat spacetimes.
\newblock {\em Phys. Rev. D}, 98:064032, Sep 2018.
\newblock URL: \url{https://link.aps.org/doi/10.1103/PhysRevD.98.064032}, \href
  {https://doi.org/10.1103/PhysRevD.98.064032}
  {\path{doi:10.1103/PhysRevD.98.064032}}.

\bibitem{Compere:2016jwb}
Geoffrey Comp{\`e}re and Jiang Long.
\newblock {Vacua of the gravitational field}.
\newblock {\em JHEP}, 07:137, 2016.
\newblock \href {http://arxiv.org/abs/1601.04958} {\path{arXiv:1601.04958}},
  \href {https://doi.org/10.1007/JHEP07(2016)137}
  {\path{doi:10.1007/JHEP07(2016)137}}.

\bibitem{Bieri:2013hqa}
Lydia Bieri and David Garfinkle.
\newblock {An electromagnetic analogue of gravitational wave memory}.
\newblock {\em Class. Quant. Grav.}, 30:195009, 2013.
\newblock \href {http://arxiv.org/abs/1307.5098} {\path{arXiv:1307.5098}},
  \href {https://doi.org/10.1088/0264-9381/30/19/195009}
  {\path{doi:10.1088/0264-9381/30/19/195009}}.

\bibitem{susskind2019electromagneticmemory}
Leonard Susskind.
\newblock Electromagnetic memory, 2019.
\newblock URL: \url{https://arxiv.org/abs/1507.02584}, \href
  {http://arxiv.org/abs/1507.02584} {\path{arXiv:1507.02584}}.

\bibitem{PhysRevLett.116.091101}
Laura Donnay, Gaston Giribet, Hern\'an~A. Gonz\'alez, and Miguel Pino.
\newblock Supertranslations and superrotations at the black hole horizon.
\newblock {\em Phys. Rev. Lett.}, 116:091101, Mar 2016.
\newblock URL: \url{https://link.aps.org/doi/10.1103/PhysRevLett.116.091101},
  \href {https://doi.org/10.1103/PhysRevLett.116.091101}
  {\path{doi:10.1103/PhysRevLett.116.091101}}.

\bibitem{PhysRevD.102.044041}
Srijit Bhattacharjee and Shailesh Kumar.
\newblock Memory effect and bms symmetries for extreme black holes.
\newblock {\em Phys. Rev. D}, 102:044041, Aug 2020.
\newblock URL: \url{https://link.aps.org/doi/10.1103/PhysRevD.102.044041},
  \href {https://doi.org/10.1103/PhysRevD.102.044041}
  {\path{doi:10.1103/PhysRevD.102.044041}}.

\bibitem{PhysRev.52.54}
F.~Bloch and A.~Nordsieck.
\newblock Note on the radiation field of the electron.
\newblock {\em Phys. Rev.}, 52:54--59, Jul 1937.
\newblock URL: \url{https://link.aps.org/doi/10.1103/PhysRev.52.54}, \href
  {https://doi.org/10.1103/PhysRev.52.54} {\path{doi:10.1103/PhysRev.52.54}}.

\bibitem{Low:1954kd}
F.~E. Low.
\newblock {Scattering of light of very low frequency by systems of spin 1/2}.
\newblock {\em Phys. Rev.}, 96:1428--1432, 1954.
\newblock \href {https://doi.org/10.1103/PhysRev.96.1428}
  {\path{doi:10.1103/PhysRev.96.1428}}.

\bibitem{Low:1958sn}
F.~E. Low.
\newblock {Bremsstrahlung of very low-energy quanta in elementary particle
  collisions}.
\newblock {\em Phys. Rev.}, 110:974--977, 1958.
\newblock \href {https://doi.org/10.1103/PhysRev.110.974}
  {\path{doi:10.1103/PhysRev.110.974}}.

\bibitem{Gell-Mann:1954wra}
Murray Gell-Mann and M.~L. Goldberger.
\newblock {Scattering of low-energy photons by particles of spin 1/2}.
\newblock {\em Phys. Rev.}, 96:1433--1438, 1954.
\newblock \href {https://doi.org/10.1103/PhysRev.96.1433}
  {\path{doi:10.1103/PhysRev.96.1433}}.

\bibitem{Yennie:1961ad}
D.~R. Yennie, Steven~C. Frautschi, and H.~Suura.
\newblock {The infrared divergence phenomena and high-energy processes}.
\newblock {\em Annals Phys.}, 13:379--452, 1961.
\newblock \href {https://doi.org/10.1016/0003-4916(61)90151-8}
  {\path{doi:10.1016/0003-4916(61)90151-8}}.

\bibitem{Weinberg:1965nx}
Steven Weinberg.
\newblock {Infrared photons and gravitons}.
\newblock {\em Phys. Rev.}, 140:B516--B524, 1965.
\newblock \href {https://doi.org/10.1103/PhysRev.140.B516}
  {\path{doi:10.1103/PhysRev.140.B516}}.

\bibitem{PhysRevD.14.2460}
S.~W. Hawking.
\newblock Breakdown of predictability in gravitational collapse.
\newblock {\em Phys. Rev. D}, 14:2460--2473, Nov 1976.
\newblock URL: \url{https://link.aps.org/doi/10.1103/PhysRevD.14.2460}, \href
  {https://doi.org/10.1103/PhysRevD.14.2460}
  {\path{doi:10.1103/PhysRevD.14.2460}}.

\bibitem{PhysRevLett.116.231301}
Stephen~W. Hawking, Malcolm~J. Perry, and Andrew Strominger.
\newblock Soft hair on black holes.
\newblock {\em Phys. Rev. Lett.}, 116:231301, Jun 2016.
\newblock URL: \url{https://link.aps.org/doi/10.1103/PhysRevLett.116.231301},
  \href {https://doi.org/10.1103/PhysRevLett.116.231301}
  {\path{doi:10.1103/PhysRevLett.116.231301}}.

\bibitem{Hawking:1975vcx}
S.~W. Hawking.
\newblock {Particle Creation by Black Holes}.
\newblock {\em Commun. Math. Phys.}, 43:199--220, 1975.
\newblock [Erratum: Commun.Math.Phys. 46, 206 (1976)].
\newblock \href {https://doi.org/10.1007/BF02345020}
  {\path{doi:10.1007/BF02345020}}.

\bibitem{Hawking:2016sgy}
Stephen~W. Hawking, Malcolm~J. Perry, and Andrew Strominger.
\newblock {Superrotation Charge and Supertranslation Hair on Black Holes}.
\newblock {\em JHEP}, 05:161, 2017.
\newblock \href {http://arxiv.org/abs/1611.09175} {\path{arXiv:1611.09175}},
  \href {https://doi.org/10.1007/JHEP05(2017)161}
  {\path{doi:10.1007/JHEP05(2017)161}}.

\bibitem{haco2018black}
Sasha Haco, Stephen~W Hawking, Malcolm~J Perry, and Andrew Strominger.
\newblock Black hole entropy and soft hair.
\newblock {\em Journal of High Energy Physics}, 2018(12):1--19, 2018.

\bibitem{PhysRevD.96.084032}
H.~Afshar, D.~Grumiller, and M.~M. Sheikh-Jabbari.
\newblock Near horizon soft hair as microstates of three dimensional black
  holes.
\newblock {\em Phys. Rev. D}, 96:084032, Oct 2017.
\newblock URL: \url{https://link.aps.org/doi/10.1103/PhysRevD.96.084032}, \href
  {https://doi.org/10.1103/PhysRevD.96.084032}
  {\path{doi:10.1103/PhysRevD.96.084032}}.

\bibitem{Chu_2018}
Chong-Sun Chu and Yoji Koyama.
\newblock Soft hair of dynamical black hole and hawking radiation.
\newblock {\em Journal of High Energy Physics}, 2018(4), April 2018.
\newblock URL: \url{http://dx.doi.org/10.1007/JHEP04(2018)056}, \href
  {https://doi.org/10.1007/jhep04(2018)056}
  {\path{doi:10.1007/jhep04(2018)056}}.

\bibitem{PhysRevD.108.044034}
Francesco Di~Filippo, Naoki Ogawa, Shinji Mukohyama, and Takahiro Waki.
\newblock Soft hair, dressed coordinates, and information loss paradox.
\newblock {\em Phys. Rev. D}, 108:044034, Aug 2023.
\newblock URL: \url{https://link.aps.org/doi/10.1103/PhysRevD.108.044034},
  \href {https://doi.org/10.1103/PhysRevD.108.044034}
  {\path{doi:10.1103/PhysRevD.108.044034}}.

\bibitem{Israel:1967za}
Werner Israel.
\newblock {Event horizons in static electrovac space-times}.
\newblock {\em Commun. Math. Phys.}, 8:245--260, 1968.
\newblock \href {https://doi.org/10.1007/BF01645859}
  {\path{doi:10.1007/BF01645859}}.

\bibitem{Israel.164.1776}
Werner Israel.
\newblock Event horizons in static vacuum space-times.
\newblock {\em Phys. Rev.}, 164:1776--1779, Dec 1967.
\newblock URL: \url{https://link.aps.org/doi/10.1103/PhysRev.164.1776}, \href
  {https://doi.org/10.1103/PhysRev.164.1776}
  {\path{doi:10.1103/PhysRev.164.1776}}.

\bibitem{PhysRevLett.26.331}
B.~Carter.
\newblock Axisymmetric black hole has only two degrees of freedom.
\newblock {\em Phys. Rev. Lett.}, 26:331--333, Feb 1971.
\newblock URL: \url{https://link.aps.org/doi/10.1103/PhysRevLett.26.331}, \href
  {https://doi.org/10.1103/PhysRevLett.26.331}
  {\path{doi:10.1103/PhysRevLett.26.331}}.

\bibitem{strominger2017blackholeinformationrevisited}
Andrew Strominger.
\newblock Black hole information revisited, 2017.
\newblock URL: \url{https://arxiv.org/abs/1706.07143}, \href
  {http://arxiv.org/abs/1706.07143} {\path{arXiv:1706.07143}}.

\bibitem{haco2019kerrnewmanblackholeentropy}
Sasha Haco, Malcolm~J. Perry, and Andrew Strominger.
\newblock Kerr-newman black hole entropy and soft hair, 2019.
\newblock URL: \url{https://arxiv.org/abs/1902.02247}, \href
  {http://arxiv.org/abs/1902.02247} {\path{arXiv:1902.02247}}.

\bibitem{PhysRevD.103.126020}
Peng Cheng and Yang An.
\newblock Soft black hole information paradox: Page curve from maxwell soft
  hair of a black hole.
\newblock {\em Phys. Rev. D}, 103:126020, Jun 2021.
\newblock URL: \url{https://link.aps.org/doi/10.1103/PhysRevD.103.126020},
  \href {https://doi.org/10.1103/PhysRevD.103.126020}
  {\path{doi:10.1103/PhysRevD.103.126020}}.

\bibitem{Donnelly:2014fua}
William Donnelly and Aron~C. Wall.
\newblock {Entanglement entropy of electromagnetic edge modes}.
\newblock {\em Phys. Rev. Lett.}, 114(11):111603, 2015.
\newblock \href {http://arxiv.org/abs/1412.1895} {\path{arXiv:1412.1895}},
  \href {https://doi.org/10.1103/PhysRevLett.114.111603}
  {\path{doi:10.1103/PhysRevLett.114.111603}}.

\bibitem{Donnelly:2015hxa}
William Donnelly and Aron~C. Wall.
\newblock {Geometric entropy and edge modes of the electromagnetic field}.
\newblock {\em Phys. Rev. D}, 94(10):104053, 2016.
\newblock \href {http://arxiv.org/abs/1506.05792} {\path{arXiv:1506.05792}},
  \href {https://doi.org/10.1103/PhysRevD.94.104053}
  {\path{doi:10.1103/PhysRevD.94.104053}}.

\bibitem{Harlow:2015lma}
Daniel Harlow.
\newblock {Wormholes, Emergent Gauge Fields, and the Weak Gravity Conjecture}.
\newblock {\em JHEP}, 01:122, 2016.
\newblock \href {http://arxiv.org/abs/1510.07911} {\path{arXiv:1510.07911}},
  \href {https://doi.org/10.1007/JHEP01(2016)122}
  {\path{doi:10.1007/JHEP01(2016)122}}.

\bibitem{Harlow:2016vwg}
Daniel Harlow.
\newblock {The Ryu{\textendash}Takayanagi Formula from Quantum Error
  Correction}.
\newblock {\em Commun. Math. Phys.}, 354(3):865--912, 2017.
\newblock \href {http://arxiv.org/abs/1607.03901} {\path{arXiv:1607.03901}},
  \href {https://doi.org/10.1007/s00220-017-2904-z}
  {\path{doi:10.1007/s00220-017-2904-z}}.

\bibitem{Maldacena:2016upp}
Juan Maldacena, Douglas Stanford, and Zhenbin Yang.
\newblock {Conformal symmetry and its breaking in two dimensional Nearly
  Anti-de-Sitter space}.
\newblock {\em PTEP}, 2016(12):12C104, 2016.
\newblock \href {http://arxiv.org/abs/1606.01857} {\path{arXiv:1606.01857}},
  \href {https://doi.org/10.1093/ptep/ptw124} {\path{doi:10.1093/ptep/ptw124}}.

\bibitem{SJFletcher_2003}
S~J Fletcher and A~W~C Lun.
\newblock The kerr spacetime in generalized bondi–sachs coordinates.
\newblock {\em Classical and Quantum Gravity}, 20(19):4153, sep 2003.
\newblock URL: \url{https://dx.doi.org/10.1088/0264-9381/20/19/302}, \href
  {https://doi.org/10.1088/0264-9381/20/19/302}
  {\path{doi:10.1088/0264-9381/20/19/302}}.

\bibitem{Barnich:2011mi}
Glenn Barnich and Cedric Troessaert.
\newblock {BMS charge algebra}.
\newblock {\em JHEP}, 12:105, 2011.
\newblock \href {http://arxiv.org/abs/1106.0213} {\path{arXiv:1106.0213}},
  \href {https://doi.org/10.1007/JHEP12(2011)105}
  {\path{doi:10.1007/JHEP12(2011)105}}.

\bibitem{Flanagan:2014kfa}
{\'E}anna~{\'E}. Flanagan and David~A. Nichols.
\newblock {Observer dependence of angular momentum in general relativity and
  its relationship to the gravitational-wave memory effect}.
\newblock {\em Phys. Rev. D}, 92(8):084057, 2015.
\newblock [Erratum: Phys.Rev.D 93, 049905 (2016)].
\newblock \href {http://arxiv.org/abs/1411.4599} {\path{arXiv:1411.4599}},
  \href {https://doi.org/10.1103/PhysRevD.92.084057}
  {\path{doi:10.1103/PhysRevD.92.084057}}.

\bibitem{Thorne:1980ru}
K.~S. Thorne.
\newblock {Multipole Expansions of Gravitational Radiation}.
\newblock {\em Rev. Mod. Phys.}, 52:299--339, 1980.
\newblock \href {https://doi.org/10.1103/RevModPhys.52.299}
  {\path{doi:10.1103/RevModPhys.52.299}}.

\bibitem{Hansen:1974zz}
R.~O. Hansen.
\newblock {Multipole moments of stationary space-times}.
\newblock {\em J. Math. Phys.}, 15:46--52, 1974.
\newblock \href {https://doi.org/10.1063/1.1666501}
  {\path{doi:10.1063/1.1666501}}.

\bibitem{Donnay_2018}
Laura Donnay, Gaston Giribet, Hernán~A. González, and Andrea Puhm.
\newblock Black hole memory effect.
\newblock {\em Physical Review D}, 98(12), December 2018.
\newblock URL: \url{http://dx.doi.org/10.1103/PhysRevD.98.124016}, \href
  {https://doi.org/10.1103/physrevd.98.124016}
  {\path{doi:10.1103/physrevd.98.124016}}.

\end{thebibliography}

\end{document}